\documentclass[a4paper,11pt]{article}
\usepackage{jcappub}
\usepackage{amsmath}
\usepackage{enumerate}
\usepackage{natbib}
\usepackage{url}
\usepackage{xcolor}
\usepackage{mathrsfs}
\usepackage{booktabs}

\newcommand{\appropto}{\mathrel{\vcenter{
  \offinterlineskip\halign{\hfil$##$\cr
    \propto\cr\noalign{\kern2pt}\sim\cr\noalign{\kern-2pt}}}}}

\title{The implications of overmassive black holes at $z > 5$ for quasar and black hole growth}

\author{Judah Luberto}

\author[]{and Steven R. Furlanetto}

\affiliation{Department of Physics and Astronomy, University of California, \\
475 Portola Plaza, Los Angeles, CA, USA}

\emailAdd{judah@astro.ucla.edu}

\abstract{
Recent JWST surveys of high-redshift galaxies have found surprisingly large black holes, with many being measured to be $\sim100$ times more massive than local galaxies with the same stellar mass. Here, we find that a population of these black holes would have dramatic implications for our understanding of their growth across cosmic time. We first show that the global black hole mass density at $z \sim 5$ would be comparable to local values. This would not occur if these black holes occupy a small fraction of galaxies, though it would be expected if these black holes radiate at high efficiencies (requiring that the central engines of AGN spin rapidly). We then show that the individual detected $z \sim 5$ black holes would remain overmassive compared to the local relation if they grow according to the average rates of state-of-the-art models. These systems must instead grow at least an order of magnitude more slowly than expected if they are to fall within the observed scatter of the local black hole mass-stellar mass relation. Such slow growth is surprising in comparison to other estimates of the radiative efficiency of AGN, especially because growth must be rapid at $z > 5$ in order to build up such massive black holes quickly. Finally, we highlight the challenges that overmassive black holes have on our understanding of the impact of quasar feedback on galaxies. 
}


\begin{document}
\maketitle
\flushbottom

\section{Introduction} \label{sec:intro}

Over the fourteen billion years from the Big Bang to today, the Universe evolved from a featureless volume to the rich, diverse environment we see at the present day.
At around cosmic noon ($z \sim 3$), many observable galaxies hosted supermassive black holes (SMBHs) at their centers. These SMBHs shape local galaxies; energetic feedback from supermassive black holes stunts the growth of nearby galaxies \cite{Kormendy&Ho2013} and is key to reproducing population statistics like the local stellar-mass function (SMF) and UV-luminosity function \cite{Somerville2008}. It is believed the opposite is true too, that host galaxies regulate the growth of SMBHs, whose stellar feedback can prevent accretion onto SMBHs (e.g., \cite{Byrne2023}). The evolution of host galaxy and SMBH are therefore linked, evidenced by tight correlations between SMBHs and hosts (see \cite{Kormendy&Ho2013} for a review). One such correlation is the $M_{\rm BH}$--$M_{*}$ relation, which compares the mass of the SMBH, $M_{\rm BH}$, to the stellar mass of the host galaxy, $M_{*}$.

This relation has been well-measured locally \cite{Reines2015_RV15, Greene2016, Bentz2018_localMstar_Mbh, Li2021_Mstar_Mbh} and is well-fit by a power law, whose parameters (power law index, normalization, and scatter) are shaped by the lives and histories of host galaxies and SMBHs. Accretion rates, mergers, environments, and feedback from the host galaxy and SMBH all feed into creating this relation
(e.g., \cite{Kauffman2000, Shankar2019_scalings, Soliman2023, Zhuang&Ho2023, Matthee2024}), and imprinted in the power law is information about the linked growth of SMBHs and host galaxies. To understand the origin of the $M_{\rm BH}$--$M_{*}$ relation, cosmologists have created models that can build the local relation from the ground up by forming early black holes (BHs) and growing them at reasonable rates until they recover the observed stellar and BH mass distribution in nearby galaxies (e.g., \cite{Lamastra2010_buildingrelation, Shankar2020, Inayoshi2022}). However, this only provides information about the \textit{local} linkage of SMBH and host galaxy because the local $M_{\rm BH}$--$M_{*}$ relation is not very sensitive to the choices of early BH growth in our models.

The first observations of $z > 4$ supermassive black holes found a population that was an order of magnitude more massive than their analogs in local galaxies at the same stellar mass \cite{Shao2017_preJWSTovermassive, Venemans2017_preJWSTovermassive, Izumi2018_preJWSTovermassive, Shimasaku2019, Wang2021_preJWSTovermassive}. However, these were measured by direct detection of their quasar light (which dominated the galaxy spectrum), so the sample was limited to only the most luminous quasars ($M_{\rm BH} \sim 10^{9}$--$10^{10} M_{\odot}$) \cite{Woo2006_bias, Volonteri2011, Onoue2019, Fan2023_review, Zhang2023_TrinityII, Tanaka2025_bias}. Being extremely high mass, it was difficult to determine whether high-$z$ SMBHs were universally ``overmassive'' or if this was a quirk of high stellar mass systems. Once the instrument sensitivities had improved, lower-mass quasars were observed ($M_{\rm BH} \sim 10^{8} M_{\odot}$) and found to fit the local relation better \cite{Willott2015, Willott2017, Izumi2019}, though the samples were still limited to relatively massive objects detected through their quasar light.

Recent observations of high-$z$ active galactic nuclei (AGN) with the \textit{James Webb Space Telescope} (JWST) have detected SMBHs with masses down to $M_{\rm BH} \sim 10^{6} M_{\odot}$ \cite{Harikane2023_AGN, Kocevski2023_littlemonsters, Larson2023, Stone2023_lowmass, Ubler2023, Bogdan2024, Greene2024, Maiolino2024_vigorous, Maiolino2024_AGN, Naidu&Matthee2024_ALT, Yue2024_AGN, Juodvbalis2025_jadesbh}. Previous surveys of SMBHs could not measure BH masses this small because they relied on direct detection of quasar light, and at lower quasar luminosities, starlight begins to outshine quasar light (see \cite{Hegde&Wyatt2024}). JWST can detect $M_{\rm BH} \sim 10^{6} M_{\odot}$ SMBHs at high redshifts indirectly through the detection of the SMBHs via rest-optical lines (e.g. Balmer lines), which are affected by dynamical broadening from emission out of the broad line region and then map their line widths to a black hole mass.

Numerous studies have found that these BHs are overmassive by $\sim 1$--$2$ orders of magnitude \cite{Pacucci2023_Mbh-Mstar, Juodvbalis2025_jadesbh, Li2025_iceberg}. This dramatic difference from the local Universe has important implications for our understanding of SMBH and galaxy evolution. So far, attention has mostly focused on building BHs this large. Achieving individual BHs at these masses is not difficult, provided BHs have higher growth efficiencies or seed masses \cite{Larson2023, Schneider2023, Volonteri2023_bias, Natarajan2024_heavyseed, Jeon2025_BHMF, Jeon2025_pathways, Sanati2025} or grow earlier \cite{Dayal2024, Escriva2024, Hooper2024, Qin2025}. Astronomers have even begun to pull physical information from the observed black hole surveys with models and have reproduced the high-$z$ $M_{\rm BH}$--$M_{*}$ relation \cite{Hu2022, Pacucci2024_connection}. 

Less attention has been focused on the implications of the overmassive $z \sim 5$ relation for black hole growth between that time and the present. In particular, if the observed high-$z$ $M_{\rm BH}$--$M_{*}$ relation is universal, are we in danger of breaking our understanding of the coevolution of SMBHs and galaxies? Are the BHs \textit{too} large to recreate local population statistics, like the local $M_{\rm BH}$--$M_{*}$ or the $z = 0$ black hole mass density? Can the high-$z$ BHs evolve across the $M_{\rm BH}$--$M_{*}$ plot to settle onto the local relation?

One crucial unknown which is necessary to answer these questions is how representative the observed overmassive BHs are of the overall galaxy population at high redshifts. The current samples require the detection of broad Balmer lines. Thus, JWST observations have so far remained limited to unobscured Type I AGN, which have broad Balmer lines, while it is thought the majority of AGN (at least at lower redshifts) are obscured Type II AGN \cite{Lusso2013_obscuredAGN}, which do not. Type II AGN are difficult to measure at high-$z$ because the narrow-line emission diagnostics used at lower redshifts make assumptions that are not applicable at high-$z$ (e.g., \cite{Harikane2023_AGN, Ubler2023, Calabro2024, Maiolino2024_AGN}). For example, the lower metallicity in a younger Universe can cause star-forming galaxies to move across the diagrams into regions AGN may occupy \cite{Feltre2016, Gutkin2016, Cameron2023, Curti2023, Sanders2023}. Various high-ionization replacement diagnostics have been proposed \cite{Brinchmann2023, Nakajima2023, Mazzolari2024, Scholtz2025}, but in the meantime the majority of the high-$z$ SMBH sample are Type I AGN.

Moreover, the observed SMBHs may not even represent the average unobscured population because of selection biases. Only luminous BHs with large emission lines are selected to be measured by spectrographs \cite{Lauer2007, Schulze2014_bias, Jeon2023_observability, Volonteri2023_bias}. While some have found that this bias cannot explain the overmassive BHs at $z \gtrsim 4$ \cite{Pacucci2023_Mbh-Mstar, Maiolino2024_AGN, Stone2024_bias}, others find the opposite, that selection bias and/or mass measurement errors \textit{can} explain the the BH measurements \cite{Li2025_iceberg, Sun2025_forwardevolution}. At slightly lower redshifts, at $z \sim 3$, there has also been no firm conclusion on the matter. Some samples do not find BHs to be overmassive up until about cosmic noon \cite{Sun2025_evolution}, while others have \cite{Mezcua2024_cosmicnoon, Luo2025}.

The abundance of massive BHs at high-$z$ is made even more confusing by a newly-discovered set of (likely) high-$z$ AGN called Little Red Dots (LRDs; e.g., \cite{Matthee2024_LRD, Labbe2025_LRD, Ma2025_LRD}) which also do not appear to follow the local $M_{\rm BH}$--$M_{*}$ relation. Even though they comprise $\sim 10\%$ of the galaxies at higher redshifts, the physical picture underlying these systems is currently unknown, due to a number of features which cannot all be explained easily. They have compact morphologies for their mass, with radii $\lesssim 50$~pc. They have broad Balmer lines, indicating AGN activity, but this can also be explained with dense, stellar systems \cite{Baggen2024_LRD_stellar, Guia2024_LRD_stellar}. They have large rest-optical emission, which might be from dust \cite{Brooks2024_yes_LRDdust}, although dust has yet to be directly detected \cite{Casey2024_LRDdust, Chen2025_LRDdust, Setton2025_LRDdust}. They have Balmer breaks, which could be from an old stellar population, but it also could arise from dense neutral hydrogen \cite{Inayoshi2025_LRD_balmer, Naidu2025_LRD_balmer}. Even the true stellar masses of LRDs are still unknown, considering they have characteristic ``V-shaped'' SEDs \cite{Ma2025_LRD_SED}, and our SED-fitting tools are not calibrated for these strange galaxies \cite{Wang2024_LRD_SED}. It is unclear where exactly the general population of LRDs place on the $M_{\rm BH}$--$M_{*}$ relation, and whether they follow a similar relation to ``normal'' galaxies, or if they follow their own scalings \cite{Graham2025_LRD_scaling}. It is uncertain if LRDs are a phase in SMBH growth or another population of galaxies entirely. These open questions only strengthen the need to characterize the growth of early SMBHs.

In this paper, we examine how the black hole population might evolve from $z \sim 5$ to the present day in light of these newly discovered populations. We find that black holes must either accrete at a very high radiative efficiency, or that the overmassive black holes must be rare. In section \ref{sec:BH_mass_density}, we consider how the overall black hole mass density grows over time. In section \ref{sec:gal_evol}, we model the growth of the observed overmassive SMBHs and test if the BHs align with the local $M_{\rm BH}$--$M_{*}$ relation when they grow at rates inferred from existing measurements and models. In section \ref{sec:max_eta_calc}, we calculate maximum BH growth rates of these galaxies to relax the tension presented in section \ref{sec:gal_evol}. Finally, in section \ref{sec:disc_and_conc}, we discuss the implications of our results and reiterate our conclusions in the paper.

Unless otherwise specified, throughout this work we use a flat $\Lambda$CDM cosmology with $\Omega_{\rm m} = 0.3111$, $\Omega_{\Lambda} = 0.6889$, $\Omega_{\rm b} = 0.0489$, $h = 0.6766$, consistent with the results of \cite{Planck2020}.

\section{Black Hole Mass Density Across Cosmic Time} \label{sec:BH_mass_density}

If high-$z$ galaxies host overmassive black holes, the global black hole mass density (BHMD) can be vastly different than previously expected. In this section, we show that the BHMD at $z \sim 5$ is nearly equal to the local BHMD if overmassive BHs are universal. This would require significant modifications to the conventional wisdom of black hole growth.

\subsection{Using the $M_{\rm BH}$--$M_{*}$ Relation to Calculate the Black Hole Mass Density} \label{sec:bhmd_from_relation}

We begin by assuming that BH masses, $M_{\rm BH}$, relate to their host galaxy's stellar masses, $M_{*}$, by a power law,

\begin{equation}
    \log M_{\rm BH} = a \log{M_{*}} + \ b, \label{eq:mbh_mstar_pl}
\end{equation}
where $a$ and $b$ are fitted coefficients. Observations of local galaxies have found that a power law provides a good fit to local data \cite{Reines2015_RV15, Greene2016, Bentz2018_localMstar_Mbh, Li2021_Mstar_Mbh}, with fitted coefficients of $a = 1.05 \pm 0.11$, $b = -4.1 \pm 0.19$ \cite{Reines2015_RV15}. We assume that the overmassive black holes occupy some fraction of galaxies, $f_{\rm over} \leq 1$; in that case, the overmassive black hole mass function can be related to the stellar mass function $\Phi(M_*,z)$ (with units number per comoving volume per unit stellar mass)
at a redshift $z$, through equation \ref{eq:mbh_mstar_pl}. The black hole mass density, $\rho_{\mathrm{BH}, z}$, for an overmassive population that fills halos in a BH mass range $(M_{\rm min},M_{\rm max})$ is then 

\begin{equation}
        \rho_{\mathrm{BH}, z} = f_{\rm over} \int_{M_{\rm min}}^{M_{\rm max}} M_{\rm BH} \Phi[M_{*}(M_{\rm BH}), z] {d M_* \over d M_{\rm BH}} dM_{\rm BH}. \label{eq:rho_bh_MbhMstar}
\end{equation}
Already we can make a simple argument why a high-$z$ BH population that is universally $\sim 100$x overmassive \cite{Harikane2023_AGN, Juodvbalis2025_jadesbh} would pose a problem. Suppose the extreme that all $z \sim 5$ galaxies align with the observed high-$z$ BHMD relation, and $f_{\rm over} = 1$ at this redshift. In this case, they all host BHs $\sim 100$x more massive than local galaxies (and scaling linearly with halo mass, as implied by measurements at both high and low redshifts \cite{Reines2015_RV15, Pacucci2023_Mbh-Mstar}). At $z \sim 5$, the stellar mass function (SMF) is $\sim 100$x less than at $z = 0$ \cite{Weaver2023_COSMO2020}. These two factors (roughly) cancel, so the BHMD at $z \sim 5$ would be approximately equal to the BHMD at $z = 0$. 

\begin{figure}
    \centering
    \includegraphics[width=0.6\linewidth]{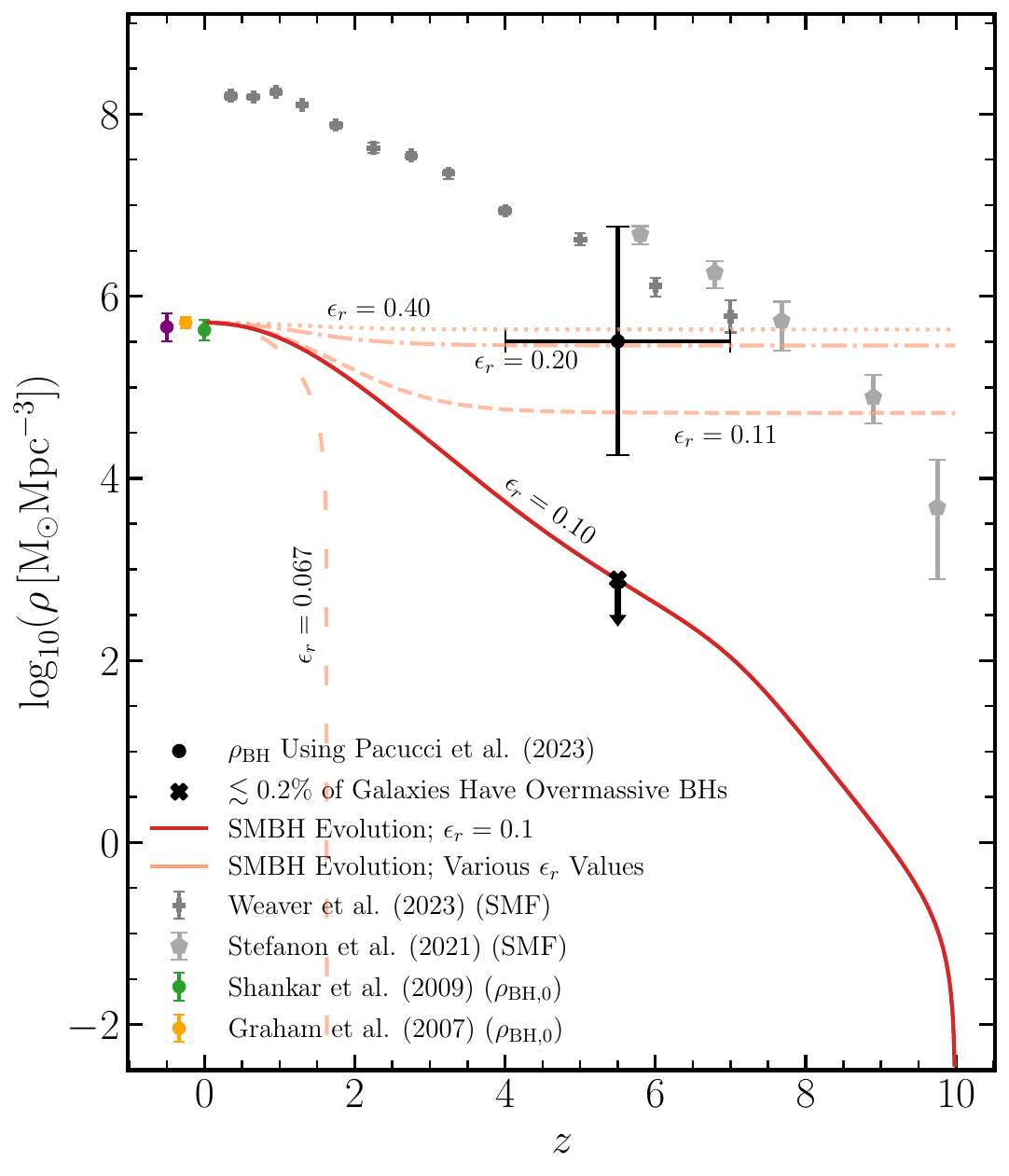}
    \caption{A demonstration of scenarios that reconcile the local and high-$z$ $M_{\rm BH}$--$M_{*}$ relations on a global level. The black point shows the black hole mass density assuming all galaxies host overmassive black holes described by the $M_{\rm BH}$--$M_{*}$ relation from \cite{Pacucci2023_Mbh-Mstar}. We use the $z \sim 6$ stellar mass function from \cite{Stefanon2021} in the conversion (larger, light gray points). The solid red line is the result of the Soltan argument using the QLF fits from \cite{Shen2020_quasar}, assuming a radiative efficiency $\epsilon_{r} = 0.1$. The light orange curves show other choices for the radiative efficiency, which depends on the spin of the BH, using the Soltan argument, including $\epsilon_{r} = 0.067$, which is the radiative efficiency estimated by \cite{Zhang2025_TrinityVI}.
    Assuming $\epsilon_r=0.1$, we scale the black point to the curve from \cite{Shen2020_quasar}, and find the fraction of galaxies hosting overmassive black holes to be $f_{\rm over} \lesssim 0.002$, or $\lesssim 0.2 \%$ of galaxies host overmassive black holes. For reference, we show a number of $z = 0$ black hole mass function measurements \cite{Marconi2004_local_rhoBH, Graham2007_local_rhoBH, Shankar2009_local_rhoBH} and stellar mass function measurements across redshift \cite{Stefanon2021, Weaver2023_COSMO2020}.}
    \label{fig:rho_bhs}
\end{figure}

For a more precise estimate, shown in figure~\ref{fig:rho_bhs}, we choose the \cite{Pacucci2023_Mbh-Mstar} fit to the $z\sim 4$--$7$ $M_{\rm BH}$--$M_{*}$ relation, who found $a=1.06 \pm 0.09$ and $b=-2.43 \pm 0.83$ (we perform the same analysis by least-squares fitting our sample in section \ref{sec:gal_evol} to find our own high-$z$ $M_{\rm BH}$--$M_{*}$ relation, with a very similar result). For the high-$z$ stellar mass function, we take the $z\sim 6$ SMF from \cite{Stefanon2021}. While this SMF predates JWST, it was already fairly well constrained at this redshift by the \textit{Hubble Space Telescope} (HST). For reference, we plot the stellar mass densities across redshifts in figure \ref{fig:rho_bhs} in gray, using both pre- and post-JWST SMF data converted to stellar mass densities \cite{Stefanon2021, Weaver2023_COSMO2020}, to show that the SMFs agree. We choose $z \sim 6$ as the redshift for our stellar mass function instead of a lower redshift in the $z \sim 4$--$7$ range of our sample because it results in a lower BHMD compared to lower redshifts. Lower redshifts have higher stellar mass density values, and therefore convert to a \textit{higher} black hole mass density value, which will only make the tension we find even stronger. We choose $[M_{\rm min},M_{\rm max}] \sim [10^{4}, 10^{12}] \ M_{\odot}$, though in section \ref{sec:calc_fover} we find that these choices do not significantly affect the result.
We plot the calculated high-$z$ black hole mass density assuming $f_{\rm over} = 1$, or all galaxies host overmassive black holes, as a black circle in figure \ref{fig:rho_bhs}, plotted at the middle of the $z \sim 4$--$7$ redshift range, $z=5.5$, with horizontal
error bars that span the redshift range and vertical error bars derived from the log-normal errors in the fitted coefficients of equation \ref{eq:mbh_mstar_pl} given by \cite{Pacucci2023_Mbh-Mstar}. The resulting BHMD ends up being very large, nearly equal to the $z = 0$ empirical measurements \cite{Marconi2004_local_rhoBH, Graham2007_local_rhoBH, Shankar2009_local_rhoBH}, which we plot in figure \ref{fig:rho_bhs} as the points near $z = 0$. (We have offset them for visual presentation, but all are actually measured locally.)

The two apparent solutions to this conundrum are that only a fraction of high-$z$ galaxies contain overmassive black holes ($f_{\rm over} \ll 1$) or that the black holes grow very little in mass from $z \sim 5$ to today. Next, we search for some insight by contextualizing the BHMD at $z \sim 5$. 

\subsection{How Much Do Black Holes Grow Through Accretion?}
\label{sec:soltan}

The Soltan argument \cite{Soltan1982} predicts the BHMD across redshift using the bolometric quasar luminosity function (QLF). If we assume a BH's bolometric luminosity, $L$, maps to the rate at which matter accretes onto the BH, $\dot{M}_{\rm acc}$, by

\begin{equation}
    L = \epsilon_{r} \dot{M}_{\rm acc} c^{2},
\end{equation}
where $\epsilon_{r} \sim 0.08$--$0.4$ is the radiative efficiency (depending on the spin rate of the BH), we can calculate a rate of accretion for a single black hole. Then, because a fraction of the accreting material is radiated away, the BH actually grows at a rate $\dot{M}_{\rm BH} = (1 - \epsilon_r) \dot{M}_{\rm acc}$.  If each black hole grows according to this rate, we can extend the calculation to a population growth rate, as long as we know the distribution of BH luminosities. This is achieved by integrating the the quasar luminosity function, $\phi(L, t)$, to find the rate of change of the BHMD:

\begin{equation}
    \mathrm{\frac{d \rho_{BH}}{dt}} = \frac{1 - \epsilon_{r}}{\epsilon_{r}c^{2}} \int_{L_{\rm min}}^{L_{\rm max}} L \phi(L, t) dL. \label{eq:soltan_result}
\end{equation}

Originally, the Soltan argument was used to integrate equation \ref{eq:soltan_result} from a very high redshift (where by assumption $\rho_{\rm BH} = 0$) to $z = 0$. With this method, local BHMD measurements are recovered as long as $\epsilon_{r} \sim 0.1$ \cite{Hopkins2007_quasarLF, Shen2020_quasar}, which has been used to help establish a (redshift independent) canonical value of $\epsilon_{r} = 0.1$. Compilations of updated BH observations have validated this canonical estimate as well. Recently, \cite{Zhang2023_trinityI, Zhang2025_TrinityVI} combined quasar luminosity functions from \cite{Shen2020_quasar} to integrate the AGN luminosity density across cosmic time and a combination of local stellar mass functions from \cite{Baldry2012_localSMF}, and the median of the local BH mass--bulge mass relations from \cite{Haring2004, Beifiori2012, Kormendy&Ho2013, McConnell2013, Savorgnan2016} to find the local black hole mass density. The comparison implied a total efficiency (a combination of radiative and kinetic efficiencies) of $\epsilon_{\rm tot} = 0.067$ using the Soltan argument.

However, in theoretical models of accretion disks the efficiency can range up to $\epsilon_{r} \approx 0.4$. Observational measurements have found a wide range of radiative efficiencies, $\epsilon_{r} \gtrsim 0.05$--$0.2$, using the spin dependence of outflows \cite{Daly2019_spin, Daly2021_spin}, X-ray continuum \cite{McClintock2006_spin, Gou2010_spin, Narzilloev2024_spin}, X-ray background \cite{Elvis2002_spin}, and clustering of AGN \cite{Shankar2020}. Simulations have measured radiative efficiencies approximately within this range as well \cite{Gruzinov1998_spin_sim, Chael2025_spin_sim}, though it can be much lower \cite{Ryan2017_spin_sim}.

Therefore, there is no special reason why the radiative efficiency should be equal to $\epsilon_{r} = 0.1$. Local BHMD measurements can still be recovered with different values of $\epsilon_{r}$ if the high-$z$ BHMD is different. To handle the uncertainty in the radiative efficiency, we integrate equation \ref{eq:soltan_result} in reverse, beginning with the local BHMD measurements and proceeding to higher redshifts. 

To perform this integral, we use the quasar luminosity functions from \cite{Shen2020_quasar}, who compiled recent ultraviolet, optical, infrared, and X-ray bands from a number of surveys and fields to create a fit to the QLF across redshift. The results are an updated version of \cite{Hopkins2007_quasarLF}, but with new data that are more robust at $z \gtrsim 3$. We plot the BHMD across cosmic time from \cite{Shen2020_quasar} using the Soltan argument and $\epsilon_{r} = 0.1$ in figure \ref{fig:rho_bhs} as a solid red line. Conveniently, this solution builds the present-day BHMD assuming that the mass density is very small at $z \sim 10$. 

However, it is clear that the inferred BHMD with $\epsilon_r=0.1$ is far smaller than the estimate at $z \sim 5$, at least using the measured relation for overmassive black holes. In the light lines in figure~\ref{fig:rho_bhs}, we use the Soltan argument by integrating backward in time to calculate the BHMD across cosmic time for various $\epsilon_{r}$ values. Because black holes grow exponentially, a larger choice of $\epsilon_{r}$ results in a much larger BHMD at high redshifts, enough that with rapidly-spinning SMBHs ($\epsilon_{r} \gtrsim 0.2$) the estimated BHMD is nearly equal to the BHMD from the high-$z$ observations! Note that the high-$\epsilon_r$ curves flatten out at high redshifts because the quasar luminosity function has fallen so much that even a tiny amount of accretion suffices to drive the AGN populations.

Thus, one way to reconcile the local BHMD, the quasar luminosity function, and estimates of BH masses at $z \sim 5$ is if nearly all accretion occurs through black holes spinning near their maximal rates. Unfortunately, this solution requires a substantial revision of our estimates of the overall radiative efficiency of AGN accretion. Of course, figure~\ref{fig:rho_bhs} also shows that the BHMD is extremely sensitive to the radiative efficiency parameter, so it is difficult to determine much more.

\subsection{Are Overmassive Black Holes Typical at High Redshifts?} \label{sec:calc_fover}

A second way to reconcile these tensions is to assume that only a fraction of the high-$z$ galaxies host overmassive black holes. Specifically, now that we can calculate the black hole mass density of the observed overmassive BHs (section \ref{sec:bhmd_from_relation}) and the universal BHMD using the Soltan argument (section \ref{sec:soltan}), we can scale the overmassive BHMD to match the Soltan argument BHMD to find a value for $f_{\rm over}$. We expect $f_{\rm over} \ll 1$ because of the simple argument in section $\ref{sec:bhmd_from_relation}$.

If $\epsilon_r \gtrsim 0.2$, we can have $f_{\rm over} \sim 1$. However, $\epsilon_{r} = 0.1$ requires $f_{\rm over} < 0.002$, or $< 0.2 \%$ of the galaxies may host overmassive BHs, which we plot in figure \ref{fig:rho_bhs} as a black cross. Therefore, there are two solutions for the overmassive BHs: either black holes are spinning nearly maximally or a small fraction of galaxies host overmassive BHs (or, a third option, that it is a mix of the two). We note that \cite{Juodvbalis2025_jadesbh} find overmassive BHs in $\sim 6\%$ of their sample, far above the required $f_{\rm over}$. Thus it is difficult to reconcile the observed samples with an efficiency $\epsilon_r=0.1$ purely through making the BHs rare.

The simplest interpretation of $f_{\rm over}$ is that a random fraction of galaxies host overmassive black holes, independent of their other properties. But another physically plausible picture is that black holes are seeded as part of the galactic evolution process, so that all galaxies above a characteristic halo mass, $M_{\rm char}$, host overmassive black holes, but smaller galaxies do not. To consider such a scenario, we can increase $M_{\rm min}$ in equation~(\ref{eq:rho_bh_MbhMstar}); we choose to use $M_{\rm min} \sim 10^{6.5} M_{\odot}$, which is the minimum black hole mass observed in the sample used in this analysis \cite{Pacucci2023_Mbh-Mstar}. Fortunately, we find that $\rho_{\mathrm{BH}, z}$ changes by $\lesssim 0.1$dex and $f_{\rm over}$ changes only marginally in this case.

With this in mind, to estimate $M_{\rm char}$, we realize in this case the overmassive fraction is equal to

\begin{equation}
    f_{\rm over} = \frac{\int_{M_{\rm char}}^{M_{\rm max}} n_h(M) dM}{\int_{M_{\rm min}}^{M_{\rm max}} n_h(M) dM}, \label{eq:min_M}
\end{equation}

where $n_h$ is the halo mass function in our redshift range, and $M_{\rm min}$ and $M_{\rm max}$ are the minimum and maximum halo masses where the halo mass function is valid. The value of $M_{\rm char}$ is not sensitive to the choice of $M_{\rm max}$, although it is sensitive to $M_{\rm min}$. Stellar masses have been found as low as $M_{*} \sim 10^{6} M_{\odot}$ at $z \sim 6$ (e.g. \cite{Chemerynska2024}), which corresponds to a minimum halo mass of $M_{h} \sim 10^{9}$ at $z \sim 6$ \cite{Behroozi2019_UniverseMACHINE}. We choose a maximum halo mass of $M_{h} \sim 10^{13}$, based on the $M_{*} \sim 10^{11} M_{\odot}$ stellar mass objects found at high-$z$ \cite{Stefanon2021}, though larger halo mass values do not change the result to $0.01$ dex. We use the halo mass function from \cite{Watson2013_hmf} for this calculation.

When solving for equation \ref{eq:min_M} (assuming $\epsilon_{r} = 0.1$), we find $\log M_{\rm char} / M_{\odot} = 10.9$ which is close to the smallest inferred halo mass hosting the observed overmassive black holes (see the next section).

\begin{table}[tbp]
\centering
\begin{tabular}{lccccr}
\toprule
    ID &  $z$ &  $\log M_{\mathrm{BH}, i} / M_{\odot}$ &  $\log M_{*, i} / M_{\odot}$ &  $\log M_{\mathrm{BH}, f} / M_{\odot}$ &  $\log M_{*, f} / M_{\odot}$ \\
\midrule

    \multicolumn{6}{c}{\textit{Juodžbalis et al. (2025)} \cite{Juodvbalis2025_jadesbh}} \\

\midrule

    GS-30148179 & 5.92 &      $7.12_{-0.35}^{+0.34}$ &        $8.78_{-0.58}^{+0.58}$ &       $9.76_{-0.21}^{+0.21}$ &        $11.02_{-0.32}^{+0.30}$ \\
    GS-10013704 & 5.92 &      $7.44_{-0.31}^{+0.31}$ &        $8.28_{-0.90}^{+0.90}$ &       $9.95_{-0.19}^{+0.19}$ &        $10.75_{-1.03}^{+0.48}$ \\
    GS-210600 & 6.31 &      $7.42_{-0.34}^{+0.33}$ &        $8.40_{-0.61}^{+0.61}$ &       $10.10_{-0.21}^{+0.21}$ &        $10.88_{-0.46}^{+0.32}$ \\
    GS-204851 & 5.48 &      $7.68_{-0.31}^{+0.32}$ &       $10.74_{-0.09}^{+0.09}$ &       $9.93_{-0.19}^{+0.20}$ &        $12.38_{-0.10}^{+0.11}$ \\
    GN-77652 & 5.23 &      $6.62_{-0.32}^{+0.33}$ &        $8.20_{-1.64}^{+1.64}$ &       $9.21_{-0.18}^{+0.19}$ &        $10.56_{-2.34}^{+0.87}$ \\
    GN-62309 & 5.17 &      $6.30_{-0.33}^{+0.32}$ &        $7.78_{-0.34}^{+0.34}$ &       $9.02_{-0.19}^{+0.20}$ &        $10.09_{-0.50}^{+0.39}$ \\
    GN-61888 & 5.87 &      $7.08_{-0.32}^{+0.34}$ &        $8.25_{-1.73}^{+1.73}$ &       $9.72_{-0.19}^{+0.20}$ &        $10.72_{-2.43}^{+0.94}$ \\
    GN-38509 & 6.68 &      $8.57_{-0.38}^{+0.36}$ &        $9.43_{-0.40}^{+0.40}$ &      $11.05_{-0.28}^{+0.31}$ &        $11.51_{-0.23}^{+0.26}$ \\
    GN-1093 & 5.59 &      $7.07_{-0.33}^{+0.33}$ &        $8.77_{-0.63}^{+0.63}$ &       $9.61_{-0.19}^{+0.20}$ &        $10.96_{-0.37}^{+0.32}$ \\
    GN-954 & 6.76 &      $7.74_{-0.32}^{+0.37}$ &        $9.68_{-0.11}^{+0.11}$ &       $10.49_{-0.21}^{+0.21}$ &         $11.68_{-0.07}^{+0.07}$ \\
    GS-172975 & 4.74 &      $7.25_{-0.32}^{+0.34}$ &        $8.98_{-0.14}^{+0.14}$ &       $9.40_{-0.18}^{+0.18}$ &        $10.94_{-0.06}^{+0.06}$ \\
    GN-73488 & 4.13 &      $7.95_{-0.30}^{+0.31}$ &        $9.71_{-0.33}^{+0.33}$ &       $9.58_{-0.17}^{+0.17}$ &        $11.16_{-0.14}^{+0.15}$ \\
    GN-53757 & 4.45 &      $7.33_{-0.31}^{+0.32}$ &       $10.38_{-0.19}^{+0.19}$ &       $9.34_{-0.17}^{+0.18}$ &        $11.56_{-0.10}^{+0.13}$ \\
    GS-38562 & 4.82 &      $7.53_{-0.31}^{+0.30}$ &        $9.76_{-0.09}^{+0.09}$ &       $9.59_{-0.18}^{+0.20}$ &        $11.31_{-0.04}^{+0.04}$ \\
    GN-20621 & 4.68 &      $7.09_{-0.34}^{+0.35}$ &        $8.41_{-1.58}^{+1.58}$ &       $9.29_{-0.19}^{+0.17}$ &        $10.62_{-2.08}^{+0.78}$ \\
    GN-11836 & 4.41 &      $7.00_{-0.32}^{+0.32}$ &        $8.17_{-0.15}^{+0.15}$ &       $9.15_{-0.17}^{+0.18}$ &        $10.35_{-0.17}^{+0.14}$ \\
    GS-8083 & 4.75 &      $7.11_{-0.31}^{+0.31}$ &        $8.27_{-0.13}^{+0.14}$ &       $9.33_{-0.17}^{+0.22}$ &        $10.52_{-0.12}^{+0.11}$ \\

\midrule

    \multicolumn{6}{c}{\textit{Harikane et al. (2023)} \cite{Harikane2023_AGN}} \\

\midrule

CEERS\_01244 & 4.48 &      $7.51_{-1.03}^{+0.63}$ &        $8.63_{-0.03}^{+0.04}$ &       $9.45_{-0.02}^{+0.02}$ &        $10.72_{-1.09}^{+0.30}$ \\
GLASS\_160133 & 4.01 &      $<6.36$ &        $8.82_{-0.02}^{+0.02}$ &       $8.67_{-0.01}^{+0.01}$ &       $<10.75$ \\
GLASS\_150029 & 4.58 &      $6.57_{-0.37}^{+0.31}$ &        $9.10_{-0.04}^{+0.02}$ &       $8.97_{-0.02}^{+0.01}$ &        $10.97_{-0.17}^{+0.14}$ \\
CEERS\_00746 & 5.62 &      $<7.76$ &        $9.11_{-0.11}^{+0.10}$ &       $10.04_{-0.07}^{+0.06}$ &        $<11.14$ \\
CEERS\_01665 & 4.48 &      $7.28_{-0.68}^{+0.51}$ &        $9.92_{-0.13}^{+0.15}$ &       $9.33_{-0.07}^{+0.08}$ &        $11.33_{-0.31}^{+0.27}$ \\
CEERS\_00672 & 5.67 &      $<7.70$ &        $9.01_{-0.13}^{+0.13}$ &       $10.01_{-0.08}^{+0.08}$ &        $<11.09$ \\
CEERS\_02782 & 5.24 &      $<7.62$ &        $9.35_{-0.12}^{+0.11}$ &       $9.80_{-0.07}^{+0.07}$ &        $<11.19$ \\
CEERS\_00397 & 6.00 &      $7.00_{-0.45}^{+0.36}$ &        $9.36_{-0.30}^{+0.26}$ &       $9.72_{-0.18}^{+0.15}$ &        $11.33_{-0.24}^{+0.20}$ \\
CEERS\_00717 & 6.94 &      $7.99_{-1.18}^{+0.77}$ &        $9.61_{-0.17}^{+0.16}$ &      $10.74_{-0.12}^{+0.12}$ &        $11.68_{-0.69}^{+1.13}$ \\
CEERS\_01236 & 4.48 &      $7.26_{-0.54}^{+0.29}$ &        $8.96_{-0.18}^{+0.19}$ &       $9.31_{-0.10}^{+0.11}$ &        $10.89_{-0.30}^{+0.13}$ \\

\midrule

    \multicolumn{6}{c}{\textit{Übler et al. (2023)} \cite{Ubler2023}}  \\

\midrule
  
    GS\_3073 & 5.55 &      $8.2_{-0.4}^{+0.4}$ &        $9.52_{-0.13}^{+0.13}$ &       $10.28_{-0.25}^{+0.26}$ &        $11.27_{-0.07}^{+0.07}$ \\
      
\bottomrule
\end{tabular}
\caption{\label{tab:sample} The IDs, redshifts, stellar masses, and black hole masses at the time of observation for our sample galaxies. Each is presented with the mass errors given from the respective sources. The $z = 0$ black hole masses and stellar masses assuming average growth (using our model) are also given, whose errors are calculated by initializing the growth of the galaxies with $\pm 1 \sigma$ for both the stellar masses and black hole masses.}
\end{table}

\section{Evolution of the $M_{\rm BH}-M_{*}$ Ratio in Individual Galaxies} \label{sec:gal_evol}

In section \ref{sec:BH_mass_density}, we showed that reconciling the overmassive BHs at $z \sim 5$ with local BHs requires either very high radiative efficiencies or (if BHs grow with the canonical radiative efficiency of 10\%) or that only a fraction $f_{\rm over} \lesssim 0.002$ of galaxies have overmassive BHs. However, we also found that there are still relatively large uncertainties in the $M_{\rm BH}-M_*$ relation, as shown in figure~\ref{fig:rho_bhs}, which translate to relatively large uncertainties in our interpretation. As a complement to these findings, in this section we explore the expected growth of a sample of observed high-$z$ galaxies hosting overmassive BHs across cosmic time and find that with average growth histories, the galaxies grow far above the local $M_{\rm BH}$--$M_{*}$  relation.

\subsection{An Overview of the AGN Sample}

We use a combination of the Type 1 AGN presented in \cite{Harikane2023_AGN, Ubler2023, Juodvbalis2025_jadesbh} for this analysis. \cite{Juodvbalis2025_jadesbh} found 34 AGN ranging from $1.5 < z < 9$, eighteen of which were $z > 4$. One of these $z > 4$ galaxies did not have a stellar mass estimate, which we removed from our sample. \cite{Harikane2023_AGN} presented ten $z > 4$ stellar mass and BH mass estimates. Three of these had stellar mass upper limits, so the propagation across the $M_{\rm BH}$--$M_{*}$ relation is also an upper limit for these relations. Finally, we took the single low-metallicity AGN from \cite{Ubler2023} at $z = 5.55$. These three samples combined to 28 different high-$z$ Type 1 AGN at $z > 4$. 

Though these samples combine into a large number of high-$z$ galaxies hosting overmassive black holes, there are a number of differences in procedure between the samples that are useful to summarize. The first difference is the selection methods, which are subtle but could lead to sampling different populations of galaxies. \cite{Harikane2023_AGN} selected AGN with broad emission in permitted lines only, so that the galaxies had broad H$\alpha$ and/or H$\beta$ with FWHM $> 1000 \rm \ km/s$, $\rm S/N > 5$ for the hydrogen lines, and other narrow forbidden optical lines [O III] and [N II] with FWHM $< 700 \ \rm km/s$. However, \cite{Juodvbalis2025_jadesbh} chose to analyze just H$\alpha$ to select galaxies (although they used other optical lines to study outflows) by splitting it into two Gaussian components and fitting the broad component to have a FWHM in the range $800 \rm \ km/s < FWHM < 10000 \ km/s$, while the narrow component had  $100 \rm \ km/s < FWHM < 800 \ km/s$. Then they repeated the same fit on H$\alpha$ for a single Gaussian component. They selected galaxies which preferred the two Gaussian component using the Bayesian information criterion (BIC), such that $\Delta \rm BIC > 5$ between the two models. Finally, \cite{Ubler2023} only had one galaxy in their sample, and had no selection criteria beyond the human component. 

The samples also differ in their methodology for black hole mass estimates. Each of their approaches is based on the method from \cite{Greene2005_BHmass}, which leverages the tight correlation between H$\alpha$ luminosity and the optical continuum strength, which is useful as the latter is a good proxy for the radius of the broad line region (a key component in estimating the mass of the black hole). The details between the papers do differ, however: \cite{Harikane2023_AGN} used the \cite{Greene2005_BHmass} relation, \cite{Juodvbalis2025_jadesbh} used \cite{Reines2015_RV15}, and \cite{Ubler2023} used \cite{Reines2013_BHmass}. While the radius-luminosity relationship between calibrations is different (both \cite{Reines2013_BHmass} and \cite{Reines2015_RV15} used the updated radius-luminosity relationship from \cite{Bentz2013}), and the datasets behind the calibrations is different as well (\cite{Reines2015_RV15} and \cite{Reines2013_BHmass} used local dwarf galaxies), we stress that these differences are not large.

The stellar masses in each dataset are computed using SED-fitting codes, of which there are a number of options. \cite{Juodvbalis2025_jadesbh} quoted values from \textsc{beagle} \cite{Chevallard2016_beagle} and \textsc{cigale} \cite{Boquien2019_cigale}, although we choose the stellar masses and errors from \textsc{beagle} because it was used with the spectroscopic data, while the fit with \textsc{cigale} used photometric data to ensure no extended stellar light was missed by the spectroscopic shutter (the stellar masses estimated with each code were within error). \cite{Harikane2023_AGN} used \textsc{prospector} \cite{Johnson2021_prospector} and \cite{Ubler2023} used \textsc{beagle} to calculate stellar mass estimates and errors.

To fit the stellar masses in a galaxy contaminated by AGN light, \cite{Harikane2023_AGN, Juodvbalis2025_jadesbh} used spectral decomposition to remove the AGN light, while \cite{Ubler2023} assumed the flux continuum was dominated by stellar light. \cite{Harikane2023_AGN} fit the high-resolution JWST and HST images with a PSF profile, a PSF and a Sérsic profile, and two PSFs and one Sérsic profile. The flux from the PSF fit(s) was removed and the resulting spectra was fit using the SED-fitting codes. Three $z > 4$ galaxies in their sample could only be well fit with the PSF profile only, while one galaxy did not have a good fit for any of the options. We report these four stellar masses as upper limits. All the other sources in their sample were best fit with one PSF and one Sérsic profile, except a single source, which was fit better with two PSFs. \cite{Juodvbalis2025_jadesbh} decomposed the AGN light from the stellar mass light by adding a power-law component to their \textsc{beagle} fits to mimic the light from an AGN and removing that component in the mass calculation. They also mask all broad lines as \textsc{beagle} does not have broad-line region models.

We take the reported stellar mass errors without modification, despite the known tendency for SED-fitting codes to underestimate the stellar mass errors, which appears to occur in the largest stellar mass systems in the sample. Fortunately, the errors would need to be greatly underestimated to alter our results substantially, and at $z \sim 5$, the biases plaguing SED-fitting codes are relatively low (within $\sim 0.5$ dex) \cite{Cochrane2025}. There is also a chance these stellar masses are uniformly biased high on account of the AGN light, though this effect would be  $\leq 0.3 \rm \ dex$ \cite{Berger2025_Ms_bias}. 

\subsection{Dark Matter Halo Growth} \label{sec:dm_growth}

In $\Lambda$CDM, the anchor and regulator of a galaxy's growth is its underlying dark matter halo. The evolution of the dark matter halo depends on mergers and accretion rates it experiences, which on short timescales can induce large fluctuations. Luckily, the growth of individual halos in $\Lambda$CDM has long been explored with simulations \cite{ShethTormen1999, Jenkins2001_HMF, Warren2006_HMF}, and on the timescale between $z \sim 5$ to $z = 0$, the fluctuations average out to be modest.

Here we assume the dark matter halos of the observed high-$z$ galaxies with overmassive BHs grow the average amount for a given mass. We take the average halo accretion rate from \cite{McBride2009_millenium}, who measured it from half a million dark matter halos in the Millennium Simulation \cite{Springel2005_millenium} (albeit with a slightly different cosmology than we use). 
They found the average to be

\begin{equation}
    \langle \dot{M} \rangle = 42 M_{\odot}/\mathrm{yr} \left( \frac{M}{10^{12} M_{\odot}}\right)^{1.127} (1 + 1.17 z) \sqrt{\Omega_{m} (1 + z)^{3} + \Omega_{\Lambda}}.
\end{equation}
We take this average relation for all of the galaxies in the sample. We do note that the \emph{average} growth rate is larger than the \emph{median} rate, because \cite{McBride2009_millenium} found that there is a long, positive tail in the $\dot{M}$ distribution of their halos. This likely reflects, for example, the environments the dark matter halos live in. This overprediction will result in more stellar mass growth in our sample, pushing the results towards the local relation (we will find the average growth histories cause our sample to lay significantly above the local relation at $z = 0$).

\begin{figure}
    \centering
    \includegraphics[width=0.5\linewidth]{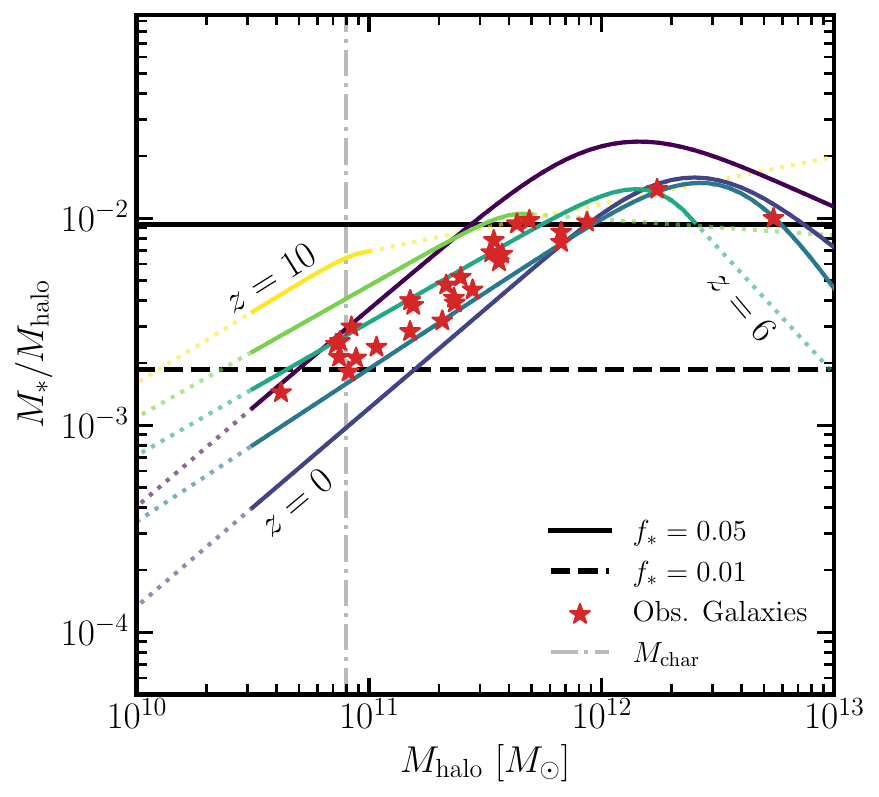}
    \caption{The connection between stellar mass and halo mass in the galaxy growth model, using results from \cite{Behroozi2019_UniverseMACHINE}. The solid colored curves are the relations from \cite{Behroozi2019_UniverseMACHINE} spanning $z = 10$ to $z = 0$, marked on the plot. The dotted curves extending from the solid lines are the linear extrapolation of the solid curves in log space.
    The conversions of the individual galaxies are marked as red stars, with most having $f_{*} = 0.01$--$0.05$. A single galaxy had a particularly high fitted stellar mass which went beyond the stellar mass range in \cite{Behroozi2019_UniverseMACHINE} for that redshift, so we set it to be $f_*=0.01$.}
    \label{fig:fstar_compare}
\end{figure}

\subsection{Halo Mass-Stellar Mass Relation} \label{sec:hmsm}

Armed with the average growth of dark matter halos in section \ref{sec:dm_growth}, we now relate the observed stellar mass of the galaxies hosting overmassive black holes to an underlying DM halo mass. This will let us estimate the average stellar mass growth rate of each galaxy.

The simplest approach is to relate the stellar mass to the halo mass using the star formation efficiency (SFE) parameter, $f_{*} = M_{*} / M_{\rm b}$. However, it is difficult to specify this parameter (although functional forms do exist, e.g. \cite{Mirocha&Furlanetto2023_bursts, Shuntov2025_SFE}), so we map the observed stellar mass to halo mass using the results from U\textsc{niverse}M\textsc{achine} \citep{Behroozi2019_UniverseMACHINE}. We also use this prescription to evolve the stellar mass from $z \sim 5$ to the present with the results from section \ref{sec:dm_growth}. We note that  the relationship derived in \cite{Behroozi2019_UniverseMACHINE} is model dependent, with most of the uncertainties at high redshifts. Fortunately, these uncertainties are much smaller than the $\sim 10$--$100$ shift required to eliminate the discrepancy, so we do not expect them to play a major role. In figure \ref{fig:fstar_compare}, we plot the $M_{*}/M_{\rm h}$ relations from \cite{Behroozi2019_UniverseMACHINE}, spanning $z=0$--$10$ in different colors, with the edge redshifts and $z = 6$ (approximately the redshift of the galaxies in our sample) labeled in the plot. We plot the relations from \cite{Behroozi2019_UniverseMACHINE} as solid lines and the linear extrapolation (in log space) to higher and lower halo masses as dotted lines.

We plot the stellar mass to halo mass ratio for our high-$z$ sample as red stars in figure \ref{fig:fstar_compare}. One of the galaxies from \cite{Juodvbalis2025_jadesbh} in our sample (at $z \sim 5.5$) had a stellar mass beyond the range outputted from \cite{Behroozi2019_UniverseMACHINE}, for which we take $M_{*}/M_{\rm h} = 0.01$ (rightmost red star in the figure). One other galaxy in the sample had an initial stellar mass of $\log M_{*}/M_{\odot} > 10$, but it was at a low enough redshift to be within the stellar mass range reported by \cite{Behroozi2019_UniverseMACHINE}. The horizontal dashed and solid lines show $f_{*} = 0.01,0.05$, respectively, which roughly bracket the star formation efficiencies in the observed sample. 
We also plot as a vertical dot-dash gray line the result to equation \ref{eq:min_M}, which is the minimum halo mass that can host overmassive black holes if all the overmassive black holes only exist in the most massive halos (see section~\ref{sec:calc_fover}). 

\begin{figure}
    \centering
    \includegraphics[width=0.5\linewidth]{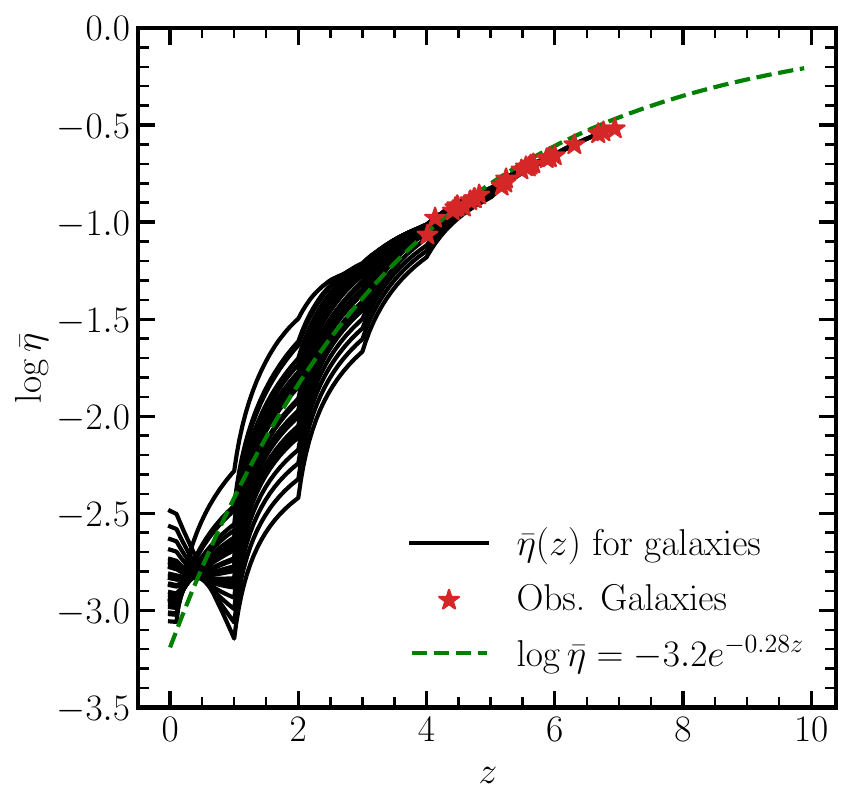}
    \caption{The Eddington ratios, $\bar{\eta}$, chosen for each of the galaxies in the sample across redshift, assuming an average BH growth rate according to \cite{Zhang2025_TrinityVI} (black curves). The initial $\bar{\eta}$ for each of the galaxies is marked as a red star. We choose the average values for $\bar{\eta}$ from \cite{Zhang2025_TrinityVI} as in this section we explore the evolution of these galaxies hosting overmassive BHs as average galaxies, although we relax the black hole mass growth rate later. We note that the data from \cite{Zhang2025_TrinityVI} are binned in $\Delta z = 1$, which we interpolated linearly in time for our calculation. When plotted in redshift space for a more convenient presentation, the curves appear to have kinks.}
    \label{fig:eta_compare}
\end{figure}

\subsection{Black Hole Growth} \label{sec:bh_growth}

We parameterize the growth of the individual black holes by assuming that black holes grow (on average) for a fraction $f_{\rm duty}$ of the available time and at a fraction $f_{\rm edd} = \dot{M}_{\rm BH}/\dot{M}_{\rm edd}$ of the Eddington limit (assuming a radiative efficiency $\epsilon_r$), where the Eddington accretion rate is $\dot{M}_{\rm edd} = (1 - \epsilon_r)/\epsilon_r \times (L_{\rm edd}/c^2)$. Here $L_{\rm edd} = 4 \pi G M_{\rm BH} m_{\rm p} c / \sigma_{\rm T}$, where $\sigma_{\rm T}$ is the Thomson cross section and $m_{\rm p}$ is the mass of a proton. We write the growth timescale as 
\begin{equation}
    \dot{M}_{\rm BH} = \bar{\eta} (t) \frac{M_{\rm BH}(t)}{t_{\rm S}} = f_{\rm edd} f_{\rm duty} {1 - \epsilon_r \over \epsilon_r} (L_{\rm edd} / c^2),
     \label{eq:eddington}
\end{equation}
where $\bar{\eta} (t) \equiv f_{\rm edd} f_{\rm duty}$ is the time-averaged Eddington ratio and the Salpeter time is $t_{\rm S} \equiv M_{\rm BH}/\dot{M}_{\rm edd} \sim 50 (\epsilon_{r} / 0.1) \ \rm Myr$. $\bar{\eta}$ is in effect an average Eddington ratio which takes into account the duty cycle of each galaxy. To model the growth of the SMBH, we must have an estimate for it.

There are many complex models which grow SMBHs by fitting their model to a number of galaxy and AGN statistics \cite{Yang2018_bhgrowth, Shankar2013, Zhang2023_trinityI}. \cite{Zhang2024_trinityIII} has calculated the redshift evolution of the quasar luminosity function using a host of population statistics (e.g., stellar mass functions, UV luminosity functions, quenched fractions), most of them pre-JWST but with the inclusion of the $9 \lesssim z \lesssim 13$ UV luminosity function from \cite{Harikane2024_UVLF}. Expanding upon \cite{Zhang2024_trinityIII}, \cite{Zhang2025_TrinityVI} has recently incorporated pre-JWST $M_{\rm BH}-M_{\rm bulge}$ and SMBH mass functions to constrain the coevolution of host galaxy and SMBH. One such parameter they calculate is the population-averaged $\bar{\eta}$ that is a function of black hole mass and redshift, which we use for evolving our sample as well.

We find $\bar{\eta}$ for each galaxy in our total sample across cosmic time, which is used to find the black hole growth for each galaxy. We plot  $\bar{\eta}$ across redshift for each galaxy in black in figure \ref{fig:eta_compare}. The modeled growth rates are large at high redshift (implying that the black holes typically accrete near Eddington for a substantial fraction of time) but decline rapidly at later times, indicating that the black holes grow much more slowly. The Eddington ratios in \cite{Zhang2025_TrinityVI} depend on redshift and black hole mass, although at high redshifts the mass dependence is weak. At $z \lesssim 3$ and $\log M_{\rm BH} / M_{\odot} > 7.5$, there is a much stronger mass dependence in the model. A few of the objects have an increasing $\bar{\eta}$ at $z < 1$, which is because in the model $\bar{\eta}$ increases a low redshifts and large black hole masses (figure 3 in \cite{Zhang2025_TrinityVI}). We note that the $\bar{\eta}$ values in this model are much lower than the results from \cite{Zhang2023_trinityI}, as it has been calibrated to a wide range of new data. Additionally,  \cite{Zhang2025_TrinityVI} finds a scatter of $\Delta \log \bar{\eta} \sim 0.3$, which is relatively small compared to, for instance, the stellar mass measurement error from our sample.

For convenience later on, we also perform a least-squares fit to the black lines to find the average growth of these BHs in the model, fitting the curves to the form

\begin{equation}
    \log \bar{\eta} (z) = \mathrm{A} e^{\mathrm{B} z}, \label{eq:eta_fit}
\end{equation}
whose fit coefficients were found to be $\rm [A,B] = [-3.2, -0.28]$. This fit is shown as the dashed green line in figure \ref{fig:eta_compare}. 

\begin{figure}
    \centering
    \includegraphics[width=\linewidth]{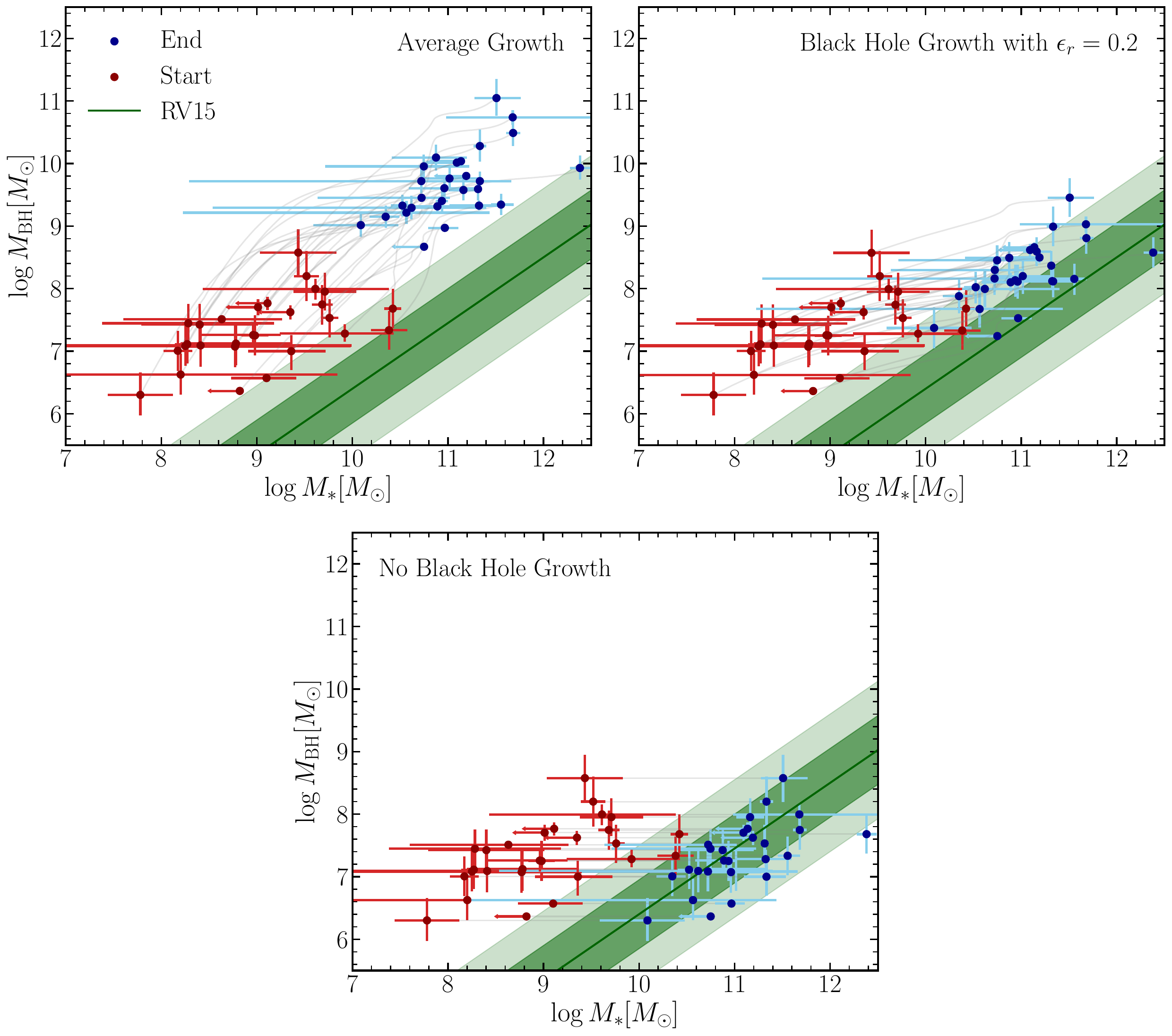}
    \caption{\textit{Top left panel:} The predicted evolution across $M_{\rm BH}$--$M_{*}$ of each galaxy using our set of models explained in section \ref{sec:gal_evol}. The red points are the starting measurements from our samples \cite{Harikane2023_AGN, Ubler2023, Juodvbalis2025_jadesbh}, with their respective errorbars (although four points from \cite{Harikane2023_AGN} are presented with stellar mass upper limits, which we show as upper limits here). The starting points evolve using our set of models across the $M_{\rm BH}$--$M_{*}$ relation in gray to the blue points, which are the predicted end values at $z = 0$, assuming average growth rates. We compare the results to the local relation \cite{Reines2015_RV15} in green, with progressively lighter green shades representing the $1$ and $2 \sigma$ scatter in the relation. The majority of the data points lie $> 3 \sigma$ away from the relation, in tension with the belief that these observed high-$z$ galaxies hosting overmassive black holes are universal. 
    \textit{Top right panel:} The same plot as the top left panel, though we assume a radiative efficiency of $\epsilon_{r} = 0.2$. The errors in the final black hole masses are calculated in the same manner as the top left panel. Here, the points remain above the local relation, but at $1$--$2\sigma$, relaxing but not resolving the tension in the top left panel. This result is interesting in context of figure \ref{fig:rho_bhs}, which found black holes radiating at a high efficiency can recover the black hole mass density at $z \sim 5$ assuming all black holes are as overmassive as recent observations suggest.
    \textit{Bottom panel:} The same plot as the top panels, except we assume there is zero black hole growth. The errors in the final black hole mass are the original measurement errors. In this case, the points settle near the local relation, and surprisingly recover (approximately) the scatter in the local relation.}
    \label{fig:growth}
\end{figure}

\subsection{Evolution of Individual Galaxies Across Cosmic Time}

We plot the results of the these growth models for our set of overmassive high-$z$ galaxies in figure \ref{fig:growth}. In the top left panel, we plot the stellar mass and BH growth of the galaxies using the models above and a radiative efficiency $\epsilon_{\rm tot}=0.067$ from \cite{Zhang2025_TrinityVI}. In the top right panel, we plot the growth of the galaxies using the same model but assuming a radiative efficiency of $\epsilon_{r} = 0.2$, and in the bottom panel we plot the growth of the galaxies assuming zero black hole growth. In each panel, the individual galaxies begin with their observed values in dark red, traverse the $M_{\rm BH}$--$M_{*}$ plot in gray, and end at $z = 0$ on the dark-blue points. We also plot the local $M_{\rm BH}$--$M_{*}$ relation from \citep{Reines2015_RV15} in dark green, with the bands showing the $1\sigma$ and $2\sigma$ scatter.

In the figure, we also account for the errors associated with the stellar mass and black hole mass observations. At the high-$z$ initial points, we use the errors reported by the observing papers. We then propagate these errors to $z=0$ by initializing the growth with $\pm 1 \sigma$ for both the stellar mass and black hole mass, and computing the result for each of the four situations at $z = 0$, represented as blue error bars. We assume the scatter in the stellar mass and black hole mass growth rates are smaller than the stellar mass and black hole mass measurement errors, so the errors in the blue points are directly from the measurement errors in our sample. However, in the bottom panel, because there is zero black hole growth, the $z=0$ black hole mass errors are the original measurement errors. 

From the top left panel, it is immediately obvious that, if the high-$z$ overmassive black holes grow at the rates inferred from independent data, they will remain overmassive at the present day, posing a substantial problem for our understanding of black hole growth. While the errors (especially in the stellar mass) can relieve some of the tension with these overmassive systems in the left panel, reconciling them would require a large systematic offset. If the problem were simply random errors in the stellar masses, we would expect that the large samples of  \cite{Harikane2023_AGN, Juodvbalis2025_jadesbh} would have found more galaxies with the same black hole masses but with stellar masses that centered on the local relation. 

The bottom panel of figure~\ref{fig:growth} shows that, if there is \emph{no} black hole growth over this time interval, the galaxies line up reasonably well with the local relation. In this case, we would conclude the stellar mass growth is well understood, but the black hole mass growth is greatly overestimated. The top right panel shows that, if $\epsilon_r=0.2$, the black holes end up somewhat above the local relation, although within the expected scatter.

\section{How Much Can Overmassive BHs Grow Before the Present Day?} \label{sec:max_eta_calc}

The top left panel of figure \ref{fig:growth} shows that, with average black hole and stellar mass growth, the overmassive systems remain $\sim 100$x above the local relation at $z = 0$. It is worrying for the entire sample to $>2$--$4\sigma$ above the local relation is a problem considering the upper $3 \sigma$ tail should only contain $\sim 0.15 \%$ of galaxies. The bottom panel shows that the simplest solution is if there is no black hole growth, in which case the galaxies nearly straddle the local relation, and may even recover its scatter, although the sample size is too small to be sure. 

In this section, we consider a middle-ground solution in which the overmassive black holes grow so as to populate the upper tail of the local $M_{\rm BH}$--$M_*$ relation. This may be possible if the observed overmassive galaxies are rare, hosting the most massive BHs for their stellar masses, because they would be responsible for only a small fraction of the black hole mass density. The galaxies could then populate only a small fraction of local galaxies. For a fixed stellar mass, this allows more black hole growth.

\subsection{Scenarios For Black Hole Rarity}
\label{sec:rarity}

To do so, we must begin with the fraction of galaxies hosting these overmassive black holes. We consider three scenarios: \emph{(a)} In section \ref{sec:BH_mass_density}, we found that these galaxies must occupy a maximum fraction of $f_{\rm over}^{a} = 0.002$ of galaxies at $z \sim 5$ for the observed BHMD to align with the prediction from the Soltan argument if $\epsilon_r=0.1$. We then assume $f_{\rm over}$ remains constant over time. \emph{(ii)} We follow \cite{Juodvbalis2025_jadesbh}, who found that their sample of overmassive black holes occupy $f_{\rm over}^b \sim 6 \%$ of galaxies at high-$z$, and again assume this fraction is constant over time. \emph{(iii)} We assume that these galaxies occupy a constant comoving density across cosmic time, rather than a constant fraction of galaxies, and therefore the fraction of galaxies that our sample occupies decreases over time. This is meant to provide a minimal estimate for the BH occupation fraction, so we begin by assuming that $f_{\rm over} \approx 0.002$ at $z=5$. Taking a minimum halo mass at that time of $\log M_{h}/M_{\odot} = 10.6$ (corresponding to the smallest systems in our sample) and comparing to the present-day halo mass function, we find that the number of halos has increased by a factor of about five, so for this scenario we set $f_{\rm over}^{c} = 0.0004$.

We now use the scatter in the local $M_{\rm BH}$--$M_{*}$ relation to determine the black hole masses to which the systems can evolve. We write the upper envelope of this as

\begin{equation}
    \log M_{\rm BH, max} = a \log{M_{*, f}} + \ b + \  k \sigma_{b}, \label{eq:Mbh_max_k}
\end{equation}
where $\sigma_{b} = 0.58$ is the scatter in the local relation \cite{Reines2015_RV15}, $M_{*, f}$ is the final stellar mass of the galaxy assuming average growth, and $k$ parameterizes the degree of scatter. If the scatter follows a gaussian distribution, $\mathscr{N}(x)$, with zero mean and standard deviation $\sigma_{b}$, then

\begin{equation}
    \int^{k \sigma_{b}}_{-k \sigma_{b}} \mathscr{N} (x) dx = 1 - 2f_{\rm over}
\end{equation}
must hold. For scenarios $a, \, b,$ and $c$, we find $k = 2.88, \, 1.55,$ and 3.35, respectively. Qualitatively, this means that if the galaxies are rarer (such as in section \ref{sec:BH_mass_density}), the individual black holes are allowed to grow more (assuming the stellar mass growth is unchanged). 

\begin{figure}
    \centering
    \includegraphics[width=0.7\linewidth]{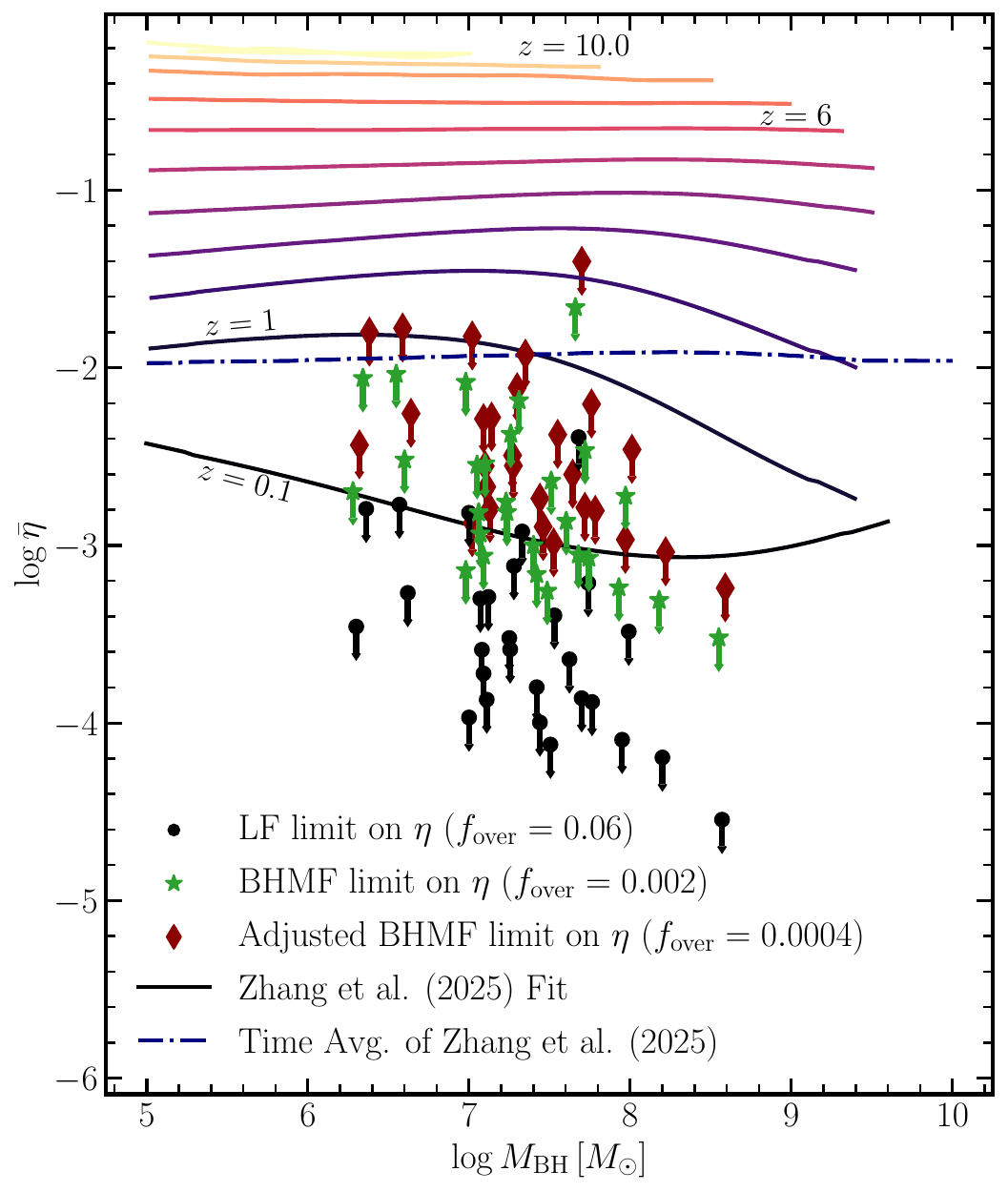}

    \caption{Estimated upper limits on the accretion efficiency, $\bar{\eta}$, for our galaxy sample to connect the observed high-$z$ overmassive black holes to the local $M_{\rm BH}$--$M_{*}$ relation, allowing for the black holes to occupy the upper tail of the local distribution. We show the limits for three scenarios (see text) with occupation fractions of the overmassive black holes, $f_{\rm over} = 0.002,\,0.06,$ and $0.0004$ as black circles, green stars, and red diamonds, respectively. Each point corresponds to one galaxy in our sample. 
    To compare these results with the average results in the \cite{Zhang2025_TrinityVI} model, we plot the average $\bar{\eta}$ as solid colored lines, spanning $z = 10$ to $z = 0$. As the upper limits on $\bar{\eta}$ are time averages, we plot as a dot-dash blue line the time average of the \cite{Zhang2025_TrinityVI} model for each initial black hole mass at $z = 5$. 
    The time average is far above the upper limit on $\bar{\eta}$ for nearly every data point, signaling that a much lower BH growth rate is needed for these overmassive BHs than models expect.}
    \label{fig:max_eta}
\end{figure}

\subsection{Estimate of the Maximum Allowed Accretion Rate}

We now estimate the maximal rate at which these black holes can grow in order to fall into the upper envelope of the observed distribution. To do so, we first assume the growth rate of our sample's black holes follows the $\bar{\eta}$ shape in equation \ref{eq:eta_fit}, which approximately fits the average BH growth in the \cite{Zhang2025_TrinityVI} model, and radiates with the same efficiency, $\epsilon_{r} = 0.067$. We have already seen that this average rate makes the black holes too large, so we fit for a new free parameter, $\log \bar{\eta} (z) = f_{\bar{\eta}} \log \bar{\eta}_{\rm fit}$. Given $\log M_{\rm BH, max}$ for each source, we estimate $f_{\bar{\eta}}$ by integrating equation \ref{eq:eddington} over redshift with the assumed $\eta(z)$, 

\begin{equation}
    f_{\bar{\eta}} = \Delta M \left[ \int^{z_{\rm obs}}_{z=0} \frac{10^{\bar{\eta}_{\rm fit} (z)}}{(1 + z) H(z)}  \frac{M_{\rm BH} (z)}{t_{S}} dz \right]^{-1}, \label{eq:f_bar_eta}
 \end{equation}
where $\Delta M = M_{\rm BH, max} - M_{\rm BH, obs}$ and $t_{\rm S}$ is the Salpeter timescale. This equation determines how much accretion is allowed, relative to the \cite{Zhang2025_TrinityVI}, for each source (given its initial mass and $M_{\rm BH, max}$). 

We numerically solve equation \ref{eq:f_bar_eta} to find a $f_{\bar{\eta}}$ for each galaxy within the three scenarios described in section \ref{sec:rarity}. We find that nearly every galaxy has $f_{\bar{\eta}} < 1$, for all three scenarios, implying that the growth must be slower than the fits from \cite{Zhang2025_TrinityVI}. To show the average trend of black hole growth in our sample, we take the average of the $f_{\bar{\eta}}$ values in our sample; for scenarios $a, \, b,$ and $c$ we find $\overline{\log f}_{\bar{\eta}} = -0.91 \pm 0.44, \, -1.70 \pm 0.49$, and $-0.64 \pm 0.44$, respectively. This means the growth of these black holes are initially growing at $\sim 1/10$ of the rate predicted in the BH growth model. 

To compare more explicitly, we use the estimated $f_{\bar{\eta}}$ values to find the time average of $\bar{\eta}$, which we  compare directly to the expectations of \cite{Zhang2025_TrinityVI} in figure \ref{fig:max_eta}. We show scenarios $a,\,b$, and $c$ (corresponding to $f_{\rm over} = 0.002,\,0.06,$ and $0.0004$) as black circles, green stars, and red diamonds, respectively. For visual purposes, the x-axis in each scenario is offset, though each scenario represents the same black hole mass. We also plot the model curves from \cite{Zhang2025_TrinityVI} as solid colored lines, ranging from $z = 10$ to $z = 0$. Nearly every upper limit sits below the $z = 1$ curve. To directly compare the model to our average $\bar{\eta}$ results, we compute the time average of the fits from \cite{Zhang2025_TrinityVI}, where we start with a range of black hole masses at $z = 5$ and grow the black holes at the average rates for each redshift. This time average is plotted as a dot-dash blue line. The dot-dash line is far above nearly every galaxy for each scenario in our sample. The conclusion is that these BHs must on average grow slowly across cosmic time (nearly at the local growth rate and $\sim 10 \%$ the average rate in our models!).

\section{Discussion and Conclusion} \label{sec:disc_and_conc}

Using a suite of simple black hole growth models, we have considered the implications of a population of $z \sim 5$ overmassive black holes and their connection to local observational statistics. We found that:

\begin{enumerate}
    \item A population of black holes at $z \sim 5$ that are universally overmassive (following the best-fit relation of \cite{Pacucci2023_Mbh-Mstar}) results in a black hole mass density that is roughly equal to the local black hole mass density (figure \ref{fig:rho_bhs}). A $z \sim 5$ black hole mass density this large can be recovered if black holes radiate at high efficiencies ($\epsilon_r > 0.2$) across cosmic time (section \ref{sec:soltan}). However, if these black holes radiate at efficiencies typical of non-rotating black holes ($\epsilon_r \approx 0.1$), which agree with current observations, the overmassive black holes at $z \sim 5$ must be exceedingly rare (section \ref{sec:BH_mass_density}) or confined only to relatively massive galaxies.

    \item If the $z \sim 5$ overmassive black holes and the galaxies hosting these overmassive black holes (see table \ref{tab:sample}) grow at average rates according to recent models (sections \ref{sec:dm_growth}, \ref{sec:hmsm}, and \ref{sec:bh_growth}), the galaxies remain $3$--$4 \sigma$ above the local $M_{\rm BH}$--$M_{*}$ relation as they evolve to $z = 0$ (figure \ref{fig:growth}; top left panel), if they have $\epsilon_r \approx 0.1$ as typically assumed. This is unexpected if the high-$z$ overmassive black holes are universal. If the overmassive black holes do not grow from $z \sim 5$ to $z = 0$, the galaxies align onto the local relation (figure \ref{fig:growth}; bottom panel). Growth at a high radiative efficiency results in black holes that are, on average, $1$--$2 \sigma$ overmassive relative to the local relation (figure \ref{fig:growth}; top right panel).

    \item If the population of galaxies hosting overmassive black holes is rare, occupying the most massive tail in the black hole mass distribution, they would be found in only a small fraction of local galaxies. Therefore, they can grow their black hole masses from $z \sim 5$ to $z = 0$ without altering the black hole mass density or the local $M_{\rm BH}$--$M_{*}$ relation. Depending on how rare these galaxies hosting overmassive black holes are at high-$z$, different rates of black holes growth are allowed (section \ref{sec:rarity}). Across a range of scenarios, the black holes must still grow at a rate only $\sim 10 \%$ of the expectations from recent models (figure \ref{fig:max_eta}), assuming that $\epsilon_r \sim 0.1$. 
\end{enumerate}

There are two potential solutions to this puzzle. In the first scenario, the overmassive black holes that have been observed so far are rare, and most galaxies have much smaller (or no) black holes at $z \sim 5$. However, these unobscured black holes have already been found in $\sim 6\%$ of galaxies at $z \sim 5$  \cite{Juodvbalis2025_jadesbh}, so unless the duty cycle of black holes is near unity at this time (in contradiction to measurements, see Fig.~\ref{fig:eta_compare}), this would be a challenge. Additionally, we have not even considered the ``little red dots'' that may also host a large population of overmassive black holes \cite{Matthee2024_LRD, Labbe2025_LRD, Ma2025_LRD}.
Nevertheless, there is substantial uncertainty about whether selection effects can explain why overmassive black holes appear universal at high redshifts.

In the second scenario, overmassive black holes are indeed common at $z \sim 5$, but at later times they accrete with very high radiative efficiencies ($\epsilon_r > 0.2$) so they grow in mass only slowly. In this case the mass density of black holes would change only by a factor of a few over the last 12~Gyr of cosmic history. The maximum radiative efficiency of a black hole (spinning at the maximum rate) is about four times larger than the canonical radiative efficiency estimated from the classic Soltan argument. However, observations using independent methods have found black holes radiate at a wide range of efficiencies, and there is no observational consensus about the average radiative efficiency, especially at high redshifts. 

Of course, a third possibility is that black hole masses at high redshifts are systematically overestimated. These masses were measured using Doppler shifting of the broad line region and the virial relation; if the local calibration of that relation does not apply at high redshifts, the black hole masses could be much smaller. This would add onto the uncertainty in the relation, which is already
large (as illustrated in figure~\ref{fig:rho_bhs}), even without accounting for any systematics. Lower mass black holes would alleviate the tension with the Soltan argument, but we note that the masses must be overestimated by orders of magnitude to remove it entirely. However, the black hole growth rates are extremely sensitive to $\epsilon_r$, so this would certainly help bring that parameter much closer to its canonical value. 

Regardless, the large samples of overmassive black holes have already changed our understanding of the population of high-redshift black holes, and it raises a number of mysteries. For example, the presence of overmassive black holes at $z > 5$ implies either that black hole seeds are extremely massive or that black hole growth is extremely efficient at high redshifts. Yet the high-$\epsilon_{r}$ solution requires that, at lower redshifts, black holes grow very \emph{inefficiently} even when they are accreting rapidly (this is the flattening in figure~\ref{fig:rho_bhs}). This would require a substantial, and fairly sudden, change in the small-scale physics of the accretion disks. Such evolution is not entirely implausible -- perhaps early super-Eddington accretion episodes ``spin up''  black holes, which then radiate efficiently because their innermost stable circular orbit is so close to the black hole -- but it requires a dramatic change in our picture of quasar evolution, which is not obviously compatible with local estimates of the efficiency parameter.

A change in accretion disk physics in supermassive black holes could also provide insight into the various ``seeding'' mechanisms which may create black holes as well. These seeds range in initial masses from stellar-mass ``light'' seeds \cite{Madau2001} to up to $M_{\mathrm{BH}, i} \sim 10^{4} M_{\odot}$ ``heavy'' seeds \cite{Loeb1994, Lodato2006, Begelman2006, Natarajan2011}, or even from more exotic seeding mechanisms like primordial black holes (up to $M_{\mathrm{BH}, i} \sim 1000 M_{\odot}$) \cite{Dayal2024, Escriva2024, Hooper2024}. Each of these seeding mechanisms require different formation times and/or accretion rates to match observations of supermassive black holes, and a sudden change in the small-scale physics of accretion disks may contain information about the presence of different seeding mechanisms.

A second puzzle suggested by these overmassive black holes
is how they affect their host galaxies, and vice versa. Black hole feedback has long been suspected to play a role in galaxy growth, as simple energetic arguments suggest (first seen in \cite{Silk&Rees1998}). Let us assume that a fraction $\epsilon_{\rm fb}$ of the accreted energy is used to unbind the gas in a galaxy, preventing further star formation. Then

\begin{equation}
    \epsilon_{\rm fb} M_{\rm BH} c^{2} \approx \frac{G M_{\rm gas} M_{h}}{R_{\rm gas}},
\end{equation}
where $R_{\rm gas} \sim \lambda R_{\rm vir}$ is the radius of the disk the gas in confined in and $\lambda$ is the spin parameter of the baryons. Taking $M_{\rm gas} = f_g (\Omega_{b}/\Omega_{m}) M_{h}$, where $f_{g}$ is the fraction of the baryonic mass that is in gas, and relating halo mass to stellar mass via the star formation efficiency, $f_{*} = M_{*} / (\Omega_b/\Omega_m \times M_{h})$, we can obtain an estimate for the $M_{\rm BH}$--$M_*$ relation under the assumption that it is set by black hole feedback. For this purpose, we use $f_{*} = 0.1 (M_{h} / 10^{12} M_{\odot})^{2/3}$ (similar to that of \cite{Behroozi2019_UniverseMACHINE}), and we find

\begin{equation}
    M_{\rm BH, fb} \sim 5 \times10^{6} M_{\odot} \left( \frac{f_{g}}{0.1} \frac{0.05}{\lambda} \frac{0.1}{\epsilon_{\rm fb}} \right) \left( \frac{M_{*}}{10^{11} M_{\odot}} \right) (1+z), \label{eq:wind}
\end{equation}
which, at $z = 0$, recovers the normalization and slope of the local $M_{\rm BH}$--$M_{*}$ relation to $\sim 0.01$--$0.1$ dex within $\log M_{*} / M_{\odot} = 7$--$12.5$. 

While the basic feedback argument of \cite{Silk&Rees1998} is certainly too simplistic to explain galaxy formation in detail, it highlights another problem with the overmassive black holes. In order to build these objects to large masses at high redshifts,
feedback must not be able to prevent nearby gas from accreting onto the black hole. Then, at later times, the
overmassive black hole must grow slowly while its surrounding stellar system grows substantially so that they begin to resemble local systems. One possible explanation for this switch would be if star formation somehow begins to prevent gas being accreted onto the central objects. However, this switch to slow black hole growth would occur during the peak of the quasar era, when we know accretion is ongoing thanks to its radiative output. According to our results, this in turn requires a high \emph{radiative} efficiency for the accretion, if overmassive black holes are common at $z \sim 5$. But, given the enormous masses of these black holes, this high radiative efficiency must be accompanied by little or no feedback on the surrounding galaxy, so that it can continue to grow! In other words, we would require $\epsilon_{\rm fb} \ll \epsilon_r$. 

A number of future observations would help settle the mysteries posed in this paper. The first is a more complete census of black hole masses at high redshifts. This would be especially helpful if galaxies with lower stellar masses and lower black hole masses are observed. The universality of these overmassive black holes is uncertain at the moment, and more observations would help settle this dispute. Further, measurements of the clustering of high-redshift sources around the overmassive black holes would allow for halo mass estimates of the sources (see \cite{CarranzaEscudero2025_LRD, Pizzati2025_LRD} for halo mass estimates of little red dots), and therefore give a better handle on how star formation has occurred in the past for these galaxies, as well as a clearer connection to the descendants of these sources at lower redshifts. Radio observations of these sources could also further constrain the inner environment of these galaxies hosting overmassive black holes, which may be leveraged to understand the feedback of these overmassive black holes. These and other future observations may help to unravel the mysteries posed by high-redshift AGNs and their connection to local galaxies and supermassive black holes. 

\acknowledgments
Judah thanks Michael Wyatt and Sahil Hegde for useful discussions. This work was supported by NASA through award 80NSSC22K0818 and by the National Science Foundation through award  AST-2205900. This work has made extensive use of NASA's Astrophysics Data System (\href{http://ui.adsabs.harvard.edu/}{http://ui.adsabs.harvard.edu/}) and the arXiv e-Print service (\href{http://arxiv.org}{http://arxiv.org}), as well as the following softwares: \textsc{matplotlib} \cite{Matplotlib}, \textsc{numpy} \cite{numpy}, \textsc{astropy} \cite{Astropy}, and \textsc{scipy} \cite{Scipy}.

\bibliography{refs.bib}{}
\bibliographystyle{JHEP.bst}

\end{document}